\documentstyle[aps,preprint,epsfig]{revtex}
\tightenlines
\title{$\Delta(1232)$ electroproduction
amplitudes in chiral soliton models of the nucleon}
\author{L.~Amoreira\footnote{amoreira@mercury.ubi.pt}}
\address{Department of Physics,
University of Beira Interior, P-6200 Covilh\~a, Portugal and
Center for Computational Physics,
University of Coimbra, P-3000 Coimbra, Portugal}
\author{P.~Alberto\footnote{pedro@teor.fis.uc.pt} \and M.~Fiolhais\footnote{tmanuel@teor.fis.uc.pt}}
\address{Department of Physics and Center for Computational
Physics, University of Coimbra, P-3000 Coimbra, Portugal}

\begin{document} 
\draft
\maketitle 
\begin{abstract} 
\noindent 
The multipole amplitudes for the N -- $\Delta$(1232) electromagnetic transition
are computed in the framework of the linear $\sigma$ model and the chiral
chromodielectric model for small and moderate photon virtualities.  The models
include quark and meson degrees of freedom and the nucleon and the delta are
clusters of three valence hedgehog quarks surrounded by meson clouds described
by coherent states.  Angular momentum and isospin projections are performed to
endow model states representing the nucleon and the delta with proper quantum
numbers.  Recoil corrections involved in the process $\gamma_{\rm v} N
\rightarrow \Delta$ are taken into account by performing linear momentum
projection of the initial and final baryon states.  The ratios $E2/M1$ and
$C2/M1$ are in good agreement with the data in the two models, but the magnetic
amplitude is better reproduced in the Linear Sigma Model.  The ratios show
little dependence with the model parameters.  Both in the Linear Sigma Model
and in the Chromodielectric Model the charged pions are responsible for the
non-vanishing quadrupole-electric and -coulomb amplitudes.  The recoil
corrections enhance the results obtained for the amplitudes without linear
momentum projection, improving the comparison with experimental data.  The
dependence of the theoretical amplitudes with the choice of the reference frame
is also studied.  
\end{abstract}


\section{Introduction}
\label{secI}
Electromagnetic processes have always played a central
role in studies of the
structure of nuclei, nucleons and their excitations.
Recently, the interest in the electroproduction of the
$\Delta$(1232) and other
nucleon resonances has increased, fueled by the large
number of experiments
planned and already running in several centers (Mainz,
Bonn, MIT, TJNAF, etc),
where very clean electromagnetic probes are now available.
In this paper we
report on a theoretical calculation of the multipole
amplitudes of the
N-$\Delta$ electromagnetic transitions.

The process $\gamma_{\rm v} N \rightarrow \Delta$
has been considered in the framework of several models of
baryon structure.
From the point of view of a pure quark model, the $\Delta$
state results from
a spin flip of one quark in the nucleon. This
corresponds to a
magnetic-dipole transition and vanishing quadrupole
transitions. The
experimental observation of quadrupole transitions, although small in
comparison with the magnetic dipole, caused a discussion
about the structure
of the nucleon and the delta. In models only with quarks,
non-vanishing
quadrupole electric and scalar nucleon-delta transition
amplitudes result from $d$-state admixtures to the quarks' lowest
$s$-state, otherwise those amplitudes would be identically zero
\cite{gogilidze,warns,capstick,bourdeau}. The quadrupole
transitions resulting
from such charge deformations are generally small. However,
other explanations can be found, in
particular the contribution of pions included in the baryon
model states
\cite{bermuth,simon,lu}, as in the type of effective theories
considered in the present work.

Our calculations are carried on in the framework of two well
known quark-meson
models of the baryon structure, namely, the Linear Sigma Model (LSM)
\cite{rlsm} and the Chiral Chromodielectric Model (CDM)
\cite{rcdm}, which
have been used to describe the structure of the nucleon
\cite{rlsm,rcdm,fiolhais,neuber,golli}.  In these models, a
baryon --- such as the
nucleon --- is a soliton with three bare valence
quarks, all in the same orbital-spin-isospin state, interacting
with chiral
$\sigma$ and $\vec \pi$ meson fields. In the CDM, there is an
additional
interaction with a scalar-isoscalar chiral singlet meson
field --- the
chromodielectric field, $\chi$. Except for a small explicit
chiral symmetry
breaking term, both models are SU(2)$\times$SU(2) chiral
invariant, a symmetry
which is spontaneously broken to SU(2), the pions being
the Goldstone bosons.

Although the two models use essentially the same ingredients,
they provide
quite different pictures of the nucleon (and of the delta).
In the LSM the stability
of the cluster of three quarks interacting with the
mesons depends on the
quark-meson interaction strength. A soliton is
formed when the coupling
constant for that interaction is sufficiently
large, and it turns out that a
strong meson cloud (particularly a pion cloud) is
required for stabilizing the
system. In the LSM,  the chiral mesons bind the three quarks.
In the CDM, besides the interaction between the quarks and
the chiral mesons, there is an interaction between the
chromodielectric field and the quarks. As a
result of this interaction the quarks acquire a
position-dependent dynamical
mass which is an increasing function of the distance.
The quarks are thus
prevented to move too far away from the origin and such
mechanism effectively
generates quark confinement. The role of the chiral mesons, although
conceptually important to implement chiral symmetry and its dynamical
breaking, gets much suppressed and the resulting picture of
the baryon is a
soliton with three confined valence quarks surrounded by a
weak cloud of
chiral mesons (particularly pions). By considering two
models providing such extreme pictures
of the nucleon, namely with quite different meson clouds,
we are able to address
interesting questions e.g. how electromagnetic transition
amplitudes are sensitive to  quantities such as the
``number of pions"
presented in the baryon states.

In the framework of the chiral models considered in
this paper, the nucleon
and the delta are made out of neutral ($\sigma$, $\chi$
and $\pi^0$) and
charged (quarks and $\pi^\pm$) particles, the latter coupling
directly to the electromagnetic probe (a virtual or a real photon).
The three quarks are assumed in the same
orbital-spin-isospin state occupying the lowest $s$-state
(no $d$-state
admixture), represented by $\left|q\right\rangle$. Moreover
the hedgehog state
is assumed for the spin-isospin wavefunction of the quarks.
The three valence
quarks are therefore described by the fully symmetric state,
$\left|q^3\right\rangle$ (anti-symmetrization
applies in color space --- the state is a Slater determinant
in color space). A
quantum mechanical description for the mesons is considered
by means of coherent states representing
pion, sigma and chi clouds, namely
$\left|\Pi\right\rangle$, $\left|\Sigma\right\rangle$ and
$\left|\chi\right\rangle$. The starting point to describe a
baryon in
the framework of the LSM and CDM is, therefore, the Fock state
$\left|\psi\right\rangle=\left|q\right\rangle^3
\left|\Pi\right\rangle
\left|\Sigma\right\rangle\left|\chi\right\rangle$.
Such state should then be projected onto angular
momentum and isospin eigenstates in order to get
states with the
nucleon and delta quantum numbers \cite{fiolhais,neuber,golli}.

The calculation reported in this work is a natural extension of
 \cite{simon}. We have refined the approximations, namely by
 taking into account
the state of motion of the initial and final baryon states
involved in the nucleon--delta transition. To this end, a
linear momentum projection \cite{lube} of the initial and
final baryon states is applied, following the
method used in~\cite{drago} for the calculation of the
nucleon electromagnetic
form factors. Besides this conceptual improvement, we
also present in more detail the  formalism and address
the issue of how the choice of the reference frame affects
the theoretical transition amplitudes.

Other calculations of the electromagnetic $N$--$\Delta$ transition
amplitudes carried out in several effective models of the nucleon have
been reported in constituent quark models
\cite{gogilidze,warns,capstick,bourdeau}, with two-body exchange
currents \cite{rth}, in the Skyrme model \cite{wirzba,walliser}, in the
cloudy bag model \cite{bermuth,lu}, in chiral quark solitons of
Nambu-Jona-Lasinio type with polarized Dirac sea \cite{watabe,silva},
etc.

This paper is organized as follows.  In Section \ref{secII} we give a
short account of the models and sketch the approximations used to
construct model states representing the nucleon and the delta.  In
Section \ref{secIII} we develop the formalism for the application of the
models to the electroproduction of the $\Delta(1232)$ with a special
emphasis on the implementation of the linear momentum projection.
Finally, the results are presented in Section \ref{secIV} together with
their discussion.  A summary of the main conclusions of this work is
presented in Section \ref{secV}.  The more technical aspects are given
in the appendices.

\section{Models and model representation of baryons}
\label{secII}

The Lagrangian densities of the LSM and the CDM can be
written, in a compact form, as
\begin{equation}
{\mathcal L} = {\mathcal L}_q + {\mathcal L}_{\sigma,\pi} +
   {\mathcal L}_{q-\sigma,\pi,\chi} + {\mathcal L}_\chi
\label{eq:1}
\end{equation}
where
\begin{eqnarray}
{\mathcal L}_q &=& i\overline{\psi}\gamma^\mu\partial_\mu\psi
\label{eq:2}\\
{\mathcal L}_{\sigma,\pi}&=&\frac{1}{2}\left(\partial\sigma\right)^2+
\frac{1}{2}\left(\partial\vec\pi\right)^2-{U}(\sigma,\vec\pi)
\label{eq:3}
\end{eqnarray}
are the pure quark and chiral meson terms,
\begin{equation}
{\mathcal L}_{q-\sigma,\pi,\chi} = \frac{g}{\chi^p}
\overline\psi\left(\sigma + {\rm i} \vec\tau\cdot\vec\pi
\gamma_5\right)\psi
\label{eq:4}
\end{equation}
is the quark-meson interaction term and
\begin{equation}
{\mathcal L}_\chi = \frac{1}{2}(\partial \chi)^2-
\frac{1}{2}M_\chi^2\chi^2,
\label{eq:5}
\end{equation}
absent in the LSM, contains the kinetic and potential terms for the
chromodielectric field. In these expressions $\psi(x)$
represents the quark
field operator, $\vec{\pi}(x)$ and $\sigma(x)$ the
chiral pion and sigma meson fields,  respectively
(the arrow denotes isovector), and $\chi(x)$ the
chromodielectric field. The parameter $p$ in the
denominator of the
interaction Lagrangian ${\mathcal L}_{q-\sigma,\pi,\chi}$
is $0$ in the LSM (no
$\chi$ field in this model) and 1 in the CDM.

The meaning of the other terms appearing in
eqs.~(\ref{eq:2})--(\ref{eq:5}) is
the following.  In Eq.~(\ref{eq:3}), ${U}(\sigma,\pi)$
is the Mexican hat
potential, $g$ in Eq.~(\ref{eq:4}) is the coupling constant which is
dimensionless in the LSM and has dimensions of energy
in the CDM (with $p=1$).
In Eq.~(\ref{eq:5}) the second term on the r.h.s. is just the
mass term for the $\chi$ field, $M_\chi$ being its mass.
Other versions of the CDM
consider a potential which includes, besides the mass
term, up to quartic terms in the $\chi$ field, as well
as other powers of $p$
in the interaction term (\ref{eq:4}). By just taking the
mass term in the potential for $\chi$ and
$p=1$ in the interaction, quark confinement is imposed in
the smoothest way,
which is the most appropriate choice for the quark matter sector of
the CDM~\cite{dragoft}.  The Mexican-hat potential is given by:
\begin{equation}
{U} = \frac{\lambda}{4}\left( \sigma^2+\vec\pi^2 - \nu^2 \right)^2 +
c\, \sigma + d.
\label{mexh}
\end{equation}
The $SU(2)\times SU(2)$ chiral symmetry of ${\mathcal L}$
is explicitly broken
by the small term $c\sigma$.  The parameters $\lambda$,
$\nu$ and $c$ are
related to the sigma and pion masses, $m_\sigma$ and $m_\pi$,
and to the pion
decay constant, $f_\pi$:
\begin{equation} \lambda= \frac{m_\sigma^2-m_\pi^2}{2 f_\pi^2}\, ,
 \ \ \ \ \ \nu^2= f_\pi^2- \frac{m_\pi^2}{\lambda}\, ,
\ \ \ \ \
c=-f_\pi\, m_\pi^2\, .
\end{equation}
In (\ref{mexh}), $d$ is a constant which guarantees
that ${\rm min }\,
{U}=0$. The Mexican hat potential induces spontaneous
chiral symmetry breaking.
The vacuum expectation values of the chiral fields are
zero for the pion and $-f_\pi$ for the sigma:
\begin{eqnarray}
\left\langle{0}\right|\hat{\vec\pi}\left|{0}\right\rangle &=& 0 \\
\left\langle{0}\right|\hat\sigma\left|{0}\right\rangle &=& -f_\pi
\label{eq:9}
\end{eqnarray}
(we use the
hat symbol ``$\hat{\mbox{\ \ }}$" whenever we want to stress the
operatorial character of the fields).
It is convenient to define a new sigma field, which we still denote by
$\hat\sigma$, as the fluctuation around the vacuum value,
i.e., we perform the
replacement $\hat\sigma \rightarrow -f_\pi + \hat\sigma$.
Hence, the vacuum expectation value of the `new' sigma
field is zero, according to (\ref{eq:9}).

Altogether, the parameters of the models defined by
(\ref{eq:1}) are the pion
and sigma masses (fixed at $m_\pi=0.139$~GeV and
$m_\sigma=1.2$~GeV), the pion
decay constant ($f_\pi=0.093$~GeV), and $g$ in the
LSM and $g$ and $M_\chi$ in the
the CDM. In the simpler version of the CDM considered
in this work, it turns out that the results are sensitive
only to
the combination $G=\sqrt{gM_\chi}$. The physical region of the
coupling constant is
$g\sim 5$ in the LSM \cite{fiolhais,golli} and $G\sim 0.2$~GeV,
in the CDM \cite{drago,dragoft}. We remark that the
physical range of the coupling constant in the CDM is much
narrower than in the LSM.  For this reason, later on, in Section IV,
when we show the dependence of the results on the coupling
constant (all other
parameters being fixed to the quoted values) we shall only consider the LSM.
The only free
parameters are $g$ in the LSM and $G$ in the CDM and these are
fixed in order to reproduce well the bulk of the nucleon properties.

It is known that, in these quark-meson models, the delta-nucleon mass
splitting is small.  We may remedy this by adding to the Hamiltonians of
the models explicit bare baryon mass terms with different masses for the
bare nucleon and the bare delta~\cite{simon,amoreira}.  Then there is
one more parameter --- the bare nucleon-bare delta mass difference ---
which can be fitted to reproduce the physical nucleon-delta mass
splitting.  The inclusion of such a term has little effect (specially in
the LSM) in the wave-functions of both quarks and mesons.  Such a bare
nucleon-delta mass splitting accounts for the residual chromomagnetic
interaction and for the 't Hooft interaction which is attractive for the
nucleon and absent in the delta.

Solutions of the LSM and CDM representing the physical
baryons can be obtained
using a variational approach based on the projected hedgehog
ansatz~\cite{neuber,golli,amoreira,rhedg}. For the sake of
completeness we sketch the
formalism here. We consider three valence quarks with spin
and isospin state
in the so-called hedgehog configuration:
\begin{equation}
\left|{hh}\right\rangle=\frac{1}{\sqrt2}\bigl(\left|{u\downarrow}
\right\rangle-\left|{d\uparrow}\right\rangle\bigr). \label{ourico}
\end{equation}
All quarks occupy the same lowest positive energy $s$-state of
the model effective potential, given by the spinor
\begin{equation}
\left\langle
     \bbox{r}\right|q_h\rangle=\frac{1}{\sqrt{4\pi}}
\left(
\matrix{
u(r)\cr
i\bbox{\sigma}\cdot\hat{\bbox{r}}v(r)\cr
}
\right)\left|{hh}\right\rangle.
\end{equation}

In our approach, the pions, sigmas and chis (in the CDM)
are described by
coherent states: $\left|{\Pi}\right\rangle$ for the pions,
$\left|{\Sigma}\right\rangle$ for the  sigmas and
$\left|{\chi}\right\rangle$ for the chis.  The expectation
values of the field operators in
these coherent states are the mean meson fields. The hedgehog
ansatz for the mesons reads:
\begin{eqnarray}
\left\langle{\Sigma}\right| \hat\sigma({\bbox r})
\left|{\Sigma}\right\rangle &=& \sigma (r) \label{mesx1} \\
\left\langle{\Pi}\right| \hat{\vec\pi} ({\bbox r})
\left|{\Pi}\right\rangle
&=& \frac{{\bbox r}}{r} \phi (r)
\label{mesx2} \\
\left\langle{\chi}\right|\hat \chi ({\bbox r})
\left|{\chi}\right\rangle
&=& \chi (r) \label{mesx3}
\end{eqnarray}
(we remember that $\hat\sigma$ is now the fluctuating
part of the original sigma
field around the vacuum expectation value, $-f_\pi$).
Actually the spherical
symmetry of sigmas and chis and the ``hedgehog-like"
character of the pion
with the peculiar isospin-coordinate space correlation,
result from  the quark
spin-isospin hedgehog configuration (\ref{ourico}) and
from the requirement of
minimum mean field energy solutions \cite{amoreira,rhedg,arriola}.

The pion coherent state is
\begin{equation}
\left|{\Pi}\right\rangle=N_\pi\left[\vec\xi\right]
\exp\left\{\sum_{i=1}^3 \int d^3\bbox k
        \sqrt{\frac{\omega_\pi(k)}{2}}\;\xi_i(\bbox k)a^\dag_i
        (\bbox k)\right\}
        \left|{0}\right\rangle
\end{equation}
where $a^\dag_i(\bbox k)$ creates a free pion with momentum
$\bbox k$ and
(Cartesian) isospin index $i$, $N_\pi$ is a normalization factor,
$\omega_\pi=\sqrt{k^2+m_\pi^2}$ and $\xi_i({\bbox k})$ is
the pion amplitude.
Similarly, the sigma coherent state is given by
\begin{equation}
\left|{\Sigma}\right\rangle=N_\sigma\left[\eta\right]\exp
\left\{ \int d^3\bbox k
        \sqrt\frac{\omega_\sigma(k)}{2}\;\eta(\bbox k) b^
        \dag(\bbox k)\right\}
        \left|{0}\right\rangle,
\end{equation}
where $b^\dag(\bbox k)$ is the sigma creation operator,
$\eta(\bbox k)$ is the
coherent state amplitude function for the $\sigma$-field.
A similar expression
holds for the chi field and we denote the amplitude of the
corresponding coherent state by
$\kappa(\bbox k)$.

The coherent states are particularly easy to deal with because they are
eigenstates of the annihilation operators, e.g.
\begin{equation}
a_i(\bbox{k})|\Pi\rangle=\sqrt{\frac{\omega_\pi(k)}{2}}\,
\xi_i(\bbox{k})\; |\Pi\rangle.
\end{equation}
Similar expressions hold,
involving the annihilation operator of sigmas and chis.

The coherent state amplitudes are the Fourier transforms
of the meson
functions in coordinate space introduced in
(\ref{mesx1})--(\ref{mesx3}), and
exhibit the following hedgehog shape in momentum space:
\begin{eqnarray}
\xi_i(\bbox k) &=& -i\, \frac{k_i}{k}\,\xi(k)\label{phedgehog}\\
\eta(\bbox k) &=& \eta(k)\\
\kappa(\bbox k) &=& \kappa(k).
\end{eqnarray}

Altogether the hedgehog baryon ansatz reads
\begin{equation}
\left|{\psi_h}\right\rangle=\left|{q_h}\right\rangle^3
                            \left|{\Sigma}\right\rangle
                            \left|{\Pi}\right\rangle
                            \left|{\chi}\right\rangle.
\label{eq:hedg}
\end{equation}
In the mean field approximation we demand the
total energy functional
$E=\left\langle{\psi_h}\right|:H:\left|{\psi_h}\right\rangle$,
where $:H:$ is the normal ordered Hamiltonian
of the models defined by (\ref{eq:1}), to be stationary
with respect to
variations of $u(r)$, $v(r)$, $\sigma(r)$, $\phi(r)$ and
$\chi(r)$. Of course, the meson
wave functions may equivalently be determined by performing
the variations
with respect to the coherent state amplitudes $\xi(k)$,
$\eta(k)$ and $\kappa(k)$. The variations with respect to
the functions of $r$ lead to a set
of differential equations.  For appropriate choices of the
coupling constants,
soliton solutions of those equations are obtained  with three quarks
absolutely confined (in CDM) \cite{rcdm,neuber} or just
bound (in LSM) \cite{rlsm,fiolhais}.

The solitons described by the hedgehog state,
$\left|{\psi_h}\right\rangle$
cannot represent physical baryons  because they are not
eigenstates of angular
momentum or isospin. In addition, (\ref{eq:hedg}) represents a
localized
object and therefore the translational symmetry of the model
Hamiltonians is
also broken in such states. In particular they contain
spurious center-of-mass
components which contribute to the energy and to other
observables.

States with good spin and isospin can be obtained from
$\left|{\psi_h}\right\rangle$ by means of the
Peierls-Yoccoz projection. The
hedgehog only contains states with $J=T$ and therefore,
due to such
space-isospace correlation, a single projection, either in
spin or in isospin,
is needed \cite{golli,rhedg}. We choose to project onto isospin.
 A baryon with isospin $T$, spin
$J=T$ and projection quantum numbers $t$ and $s$
(for isospin and spin, respectively) is given by
\begin{eqnarray}
\left|{T,\, t;J=T,\,s}\right\rangle&=& (-1)^{T+t}
{\mathcal P}^T_{t\,-s} \left|{\psi_h}\right\rangle,
\label{eq:e10}
\end{eqnarray}
where ${\mathcal P}^T_{t\,-s}$ is the
isospin operator
\begin{equation}
{\mathcal P}^T_{t\,s}=
\frac{2T+1}{8\pi^2}\int d\Omega \, {\mathcal D}_{ts}^{T^*}(\Omega)
 R(\Omega).
\end{equation}
In this expression, $R(\Omega)$ stands for the rotation
operator in isospin space, ${\mathcal D}$ are
the Wigner matrices and the
integration is performed over all orientations
$\Omega$ (which represents the
set of three Euler angles in isospin space). In
the following we consider $s=-t=M$ and use the
shorthand notation $P_{JM}={\mathcal P}^T_{M,M}$.

On the other hand, a model state representing a
baryon at rest can be obtained by applying a
Peierls-Yoccoz projection onto linear momentum zero
to the state already
projected onto isospin (and angular momentum). The
Peierls-Yoccoz linear
momentum projector is given by
\begin{equation}
P_{\bbox q}=
\frac{1}{ (2\pi)^3 } \int d {\bbox a}\,  {\rm e}^{{\rm i}
{\bbox a}\cdot{\bbox q}}
U ({\bbox a}),
\end{equation}
where $U({\bbox a})$ is the translation operator. A nucleon
at rest is
therefore represented by the model state
\begin{equation}
\left|{J=T=\frac{1}{2}, M, {\bbox q}=0}\right\rangle =
P_{{\bbox q}=0} P_{JM}
\left|{\psi_h}\right\rangle\, =
P_{{\bbox q}=0}  \left|{\psi_{JM}}\right\rangle\,.
\label{estadoq0}
\end{equation}
For ${\bbox q}=0$, the isospin-angular momentum projector
operator and the
linear momentum projector operator commute, but this is no
longer the case
for ${\bbox q}\not= 0$ \cite{neuberg}.

In order to include recoil effects in the calculations, in
principle one should boost
\cite{lube} the zero momentum states (\ref{estadoq0}), but
the technical
difficulties associated with boosting prevent, in practice,
the use of such a
procedure. However, at least for small linear momentum
$\bbox q$, we may
approximate the boost operation by the Peierls-Yoccoz
projection  onto
linear momentum $\bbox q$ \cite{neuber,neuberg}.  Thus,
the model state representing a physical
baryon of angular momentum and isospin $J$ and linear momentum
$\bbox q$ is
\begin{equation}
\left|{J=T, M, {\bbox q}}\right\rangle \sim P_{{\bbox q}}
\left|{\psi_{JM}}\right\rangle\,.
\end{equation}
Proper normalization of the projected state requires the inclusion of
kinematical normalization factors. For example a nucleon
with four-momentum
$q$, $\left|{N(q)}\right\rangle$, is described by
\begin{equation}
\left|{N(q)}\right\rangle \rightarrow \sqrt{(2 \pi)^3 \,
\delta^3(0)} \sqrt{\frac{E}{m_N}}
\frac{ P_{\bbox q}  \left|{\psi_{JM}}\right\rangle}
{\sqrt{ \left\langle{P_{\bbox q} \psi_{JM} }\right|
\left.{P_{\bbox q} \psi_{JM}}\right\rangle   }}  ,
\label{estadoq}
\end{equation}
where $J=\frac{1}{2}$, $m_N$ is the nucleon mass and $E=\sqrt{q^2+m^2_N}$ its energy.

Before presenting the formalism to compute the amplitudes for the
electroproduction of the delta, one should briefly mention
how the radial profiles $u(r)$, $v(r)$,
$\sigma(r)$, $\phi(r)$ and $\chi(r)$ are determined. They
may be determined in the so-called
``variation-before-projection" (VBP) method, and, in that case,
 the stationarity of the mean field energy is required. A better
approach (even if much more demanding numerically)
is the VAP (``variation-after-projection") method, where the energy functional
to be minimized
is the expectation value of the normal ordered Hamiltonian
in the projected
state $\left|{\psi_{JM}}\right\rangle$.
In this procedure, which we followed, one obtains
different field radial
profiles $u_B(r)$, $v_B(r)$, $\phi_B(r)$, $\sigma_B(r)$,
$\chi_B(r)$ (and
coherent state amplitudes $\xi_B(k),$ $\eta_B(k),$ $\kappa_B(k)$)
for the nucleon
($B\rightarrow N$) and for the delta ($B\rightarrow\Delta$).
The results presented Section \ref{secIV} use the VAP method for the
angular momentum projection and the approximate VAP method for
the linear momentum projection as described in \cite{neuber}.
Unless otherwise stated, the coupling constants are $g=5$
in the LSM \cite{rlsm,golli,rpa} and $G=0.2$~GeV in
the CDM~\cite{drago,dragoft}, for  which nucleon properties
are well described. These values, together with the above mentioned
values for meson masses and pion decay constant,
will be referred to as the
{\em standard} parameters.

\section{Multipole amplitudes}
\label{secIII}

The N-$\Delta$ electromagnetic transverse helicity
amplitudes~\cite{donn} are defined by
\begin{equation}
A_\lambda^{(\mu)}= -\frac{e}{\sqrt{2k_W}}
\left\langle{\Delta\;{\scriptsize{\frac{1}{2}}}\,
\lambda;\bbox{k}_\Delta}\right|
{:\bbox{\epsilon}_{\mu}}\cdot{\bbox{J}(0)}:
\left|{N\;{\scriptsize{\frac{1}{2}}}\,\lambda-\mu;
\bbox{k}_N}\right\rangle,
\label{eqe28}
\end{equation}
and the scalar helicity amplitude by
\begin{equation}
S_\lambda=\frac{1}{\sqrt{2}}\frac{e}{\sqrt{2k_W}}
\left\langle{\Delta\;{\scriptsize{\frac{1}{2}}}\,\lambda;
\bbox{k}_\Delta}\right|
:J^0(0):
\left|{N\;{\scriptsize{\frac{1}{2}}}\,\lambda;\bbox{k}_N}
\right\rangle,
\label{eqe29}
\end{equation}
where $J^\mu$ is the electromagnetic current density operator,
$\bbox{\epsilon}_i,$ $i=0,\pm 1$ are the photon polarization vectors
($\bbox{\epsilon}_0$ is chosen along the direction of the photon motion), and
$k_W$ is the magnitude of the photon three-momentum at the photon point
\cite{bourdeau}. Because of gauge invariance,
the other amplitude --- longitudinal amplitude --- is just the scalar
amplitude multiplied
by the kinematical factor $\omega/k$.
The values
for $\lambda$ and $\mu$ are usually chosen as $\lambda=1/2,\;
3/2,$ $\mu=1$ for
the transverse amplitudes and $\lambda=1/2$ for the scalar amplitude.
If the linear momentum projection of the model states is
skipped (i.e. no
recoil corrections), expressions (\ref{eqe28}) and (\ref{eqe29})
reduce to those usually presented in the literature [see e.g. Eq.
(7) of ref.~\cite{simon}]
 using the procedure described in~\cite{rpa}.

Replacing the baryon states above by their model representations in
Eq.~(\ref{estadoq}), and noting that
${U}^\dag(\bbox{r}) J^{\nu}(0){U}(\bbox{r})=J^{\nu}(\bbox{r})$,
we get (for $\lambda=1/2,$ $\mu=1$)
\begin{eqnarray}
A_{1/2}&=&
-N_{N\Delta}
\int d^3\bbox{a}d^3\bbox{r} e^{-i\bbox{q}\cdot\bbox{r}}
\left\langle \Delta\frac{1}{2}\frac{1}{2}\right|
{U}^\dag\bigl[(x-1)\bbox a\bigr]
\;:\bbox\epsilon_1\cdot\bbox{J}(\bbox{r}): \;
{U}\bigl(x\bbox{a}\bigr)
\left| N\frac{1}{2}\,-\frac{1}{2}\right\rangle
\label{atransv}\\
S_{1/2}&=&
\frac{1}{\sqrt2} N_{N\Delta}
\int d^3\bbox{a}d^3\bbox{r} e^{-i\bbox{q}\cdot\bbox{r}}
\left\langle \Delta\frac{1}{2}\frac{1}{2}\right|
{U}^\dag\bigl[(x-1)\bbox a\bigr]
\;:J^0(\bbox r):\;
{U}\bigl(x\bbox{a}\bigr)
\left| N\frac{1}{2}\frac{1}{2}\right\rangle,\label{ascalar}
\end{eqnarray}
where $x$ is the fraction of the photon momentum  carried by the
delta. In this way, the parameter $x$ identifies the reference
frame used in the calculations:
$x=0$ corresponds to the delta rest frame which is mostly used
in the literature.
The factor $N_{N\Delta}$ contains all  kinematical factors as
well as the
projected states normalization terms [see (\ref{estadoq})],
and is given by
\begin{equation}
N_{N\Delta}=
\frac{1}{(2\pi)^3}\frac{e}{\sqrt{2k_W}}
\sqrt{\frac{E_N}{m_N}}\sqrt{\frac{E_\Delta}{m_\Delta}}
\frac{1}{\sqrt{F_{1/2}\bigl[(x-1)\bbox q\bigr]F_{3/2}
\bigl(x\bbox q\bigr)}},
\end{equation}
with
\begin{equation}
F_{1/2}({q})=
\left\langle N\frac{1}{2}\frac{1}{2}
        \right| P_{\bbox{q}} \left|
N\frac{1}{2}\frac{1}{2}\right\rangle,
\end{equation}
and similarly for the delta $F$ factor. These factors only depend on the
magnitude $|\bbox{q}|$.

The multipole $N-\Delta$ transition amplitudes are
usually extracted from the helicity amplitudes above
making a multipole
expansion of the electromagnetic field. For nucleon
and delta model states which are
eigenstates of the angular momentum and parity operators,
only the magnetic
dipole and the electric and scalar quadrupole terms contribute to the
transition (see Appendix~\ref{appa} for details).
The linear momentum projection in our approach affects the
rotational symmetry of the baryon states,  and the relevant
multipoles are not automatically selected.
Instead, one has to explicitly remove the spurious terms
in the multipole
expansion of the exponential in equations (\ref{atransv})
and (\ref{ascalar}),
which implies restricting the
momentum transfer $q$ to low values. In that case,
the rotational symmetry of the model states is
almost preserved even if the linear momentum projection is performed.
The multipole amplitudes are then
\begin{eqnarray}
M^{{M1}}(q) &=&-\frac{3}{2} N_{N\Delta} \int d^3\bbox{a}\,
d^3\bbox{r}\; j_1(qr)\nonumber\\
&&
\left\langle{\Delta\;{\scriptsize{\frac{1}{2}}}\,
{\scriptsize{\frac{1}{2}}}}\right|
{U}^\dag\bigl[(x-1)\bbox{a}\bigr]
:\bigl[\hat{\bbox{r}}\times\bbox{J}(\bbox{r})\bigr]_1:
\;{U}\bigl(x\bbox{a}\bigr)
\left|{N\;{\scriptsize{\frac{1}{2}}}\,-{\scriptsize{\frac{1}{2}}}}
\right\rangle
\label{m1def}
\\
M^{{E2}}(q)&=&-\frac{\sqrt{10\pi}}{k}N_{N\Delta}
\int d^3\bbox{a}\,d^3\bbox{r}\nonumber\\
&&
\left\langle{\Delta\;{\scriptsize{\frac{1}{2}}}\,
{\scriptsize{\frac{1}{2}}}}\right|{U}^\dag
\bigl[(x-1)\bbox{a}\bigr]:
\bigl[
\bbox{\nabla}\times j_2(qr)\bbox{Y}^1_{22}(\hat{\bbox{r}})
\bigr]\cdot\bbox{J}(\bbox{r}):{U}\bigl(x\bbox{a}\bigr)
\left|{N\;{\scriptsize{\frac{1}{2}}}\,-{\scriptsize{\frac{1}{2}}}}
\right\rangle
\label{e2def}
\end{eqnarray}
\begin{eqnarray}
M^{{C2}}(q) &=&-\sqrt{10\pi}\,N_{N\Delta}\int d^3\bbox{a}\,
d^3\bbox{r}\,
j_2(qr)Y_{20}(\hat{\bbox{r}})\nonumber\\
&&
\left\langle{\Delta\;{\scriptsize{\frac{1}{2}}}\,{\scriptsize
{\frac{1}{2}}}}
\right|
{U}^\dag\bigl[(x-1)\bbox{a}\bigr]
:J^0(\bbox{r}):
\;{U}\bigl(x\bbox{a}\bigr)
\left|{N\;{\scriptsize{\frac{1}{2}}}\,{\scriptsize{\frac{1}{2}}}}
\right\rangle,
\label{c2def}
\end{eqnarray}
where $\bbox{Y}^m_{Jl}$ are the vector spherical harmonics, $j_l(x)$
are the
spherical Bessel functions and the index 1 in the $M1$ operator denotes
 component $+1$ in the spherical basis.
It is worth noticing that formulas~(\ref{m1def})--(\ref{c2def}) differ
from those used when no recoil corrections are considered (see e.g.
eqs. (10)--(12) of \cite{simon}) by the integration
over $\bbox{a}$ and by the presence of the translation operations.
Had we inserted $\delta^3(\bbox{a})$ in (\ref{m1def})--(\ref{c2def})
and integrated over $\bbox{a}$, the
expressions for the multipole amplitudes when no recoil effects are
considered (see, amongst others, \cite{simon,wirzba,watabe,silva})
would be obtained.

The electric quadrupole amplitude involves
the operator
\begin{equation}
\hat{O}_{\rm E2}(q)=\frac{1}{q}\int d^3\bbox{r}
\bigl[ \bbox \nabla\times
j_2(qr)\bbox{Y}^1_{22}(\hat{\bbox{r}})\bigr]\cdot\bbox{J}(\bbox{r}),
\end{equation}
which, using the properties of the vector spherical harmonics and
integration
by parts, can be written as
\begin{eqnarray}
\hat{O}_{\rm E2}(q)&=&\frac{1}{\sqrt 6}\;\frac{\omega}{q} \int
d^3\bbox{r}\,
     \frac{d}{dr}\bigl[r j_2(qr)\bigr]Y_{21}(\hat{\bbox r})
     J^{0}(\bbox r)
        \nonumber\\
     &&-\frac{iq}{\sqrt 6} \int d^3\bbox r\,j_2(qr)\,Y_{21}
     (\hat{\bbox r})\,
        \bbox r\cdot\bbox{J}(\bbox r),
\end{eqnarray}
where we used the electric current conservation condition to
simplify the 
first term. The second term gives a negligible correction to
the $E2$ amplitude in the low momentum regime and can be dropped
\cite{simon}. Other technical aspects of
the calculation of the multipole amplitudes are provided in
the appendices~\ref{ch:trov}~and~\ref{appb}.

\section{Results and discussion}
\label{secIV}

The ratios $E2/M1$ and $C2/M1$ for the delta electroproduction
are related to the multipoles (\ref{m1def})--(\ref{c2def}) through
\begin{eqnarray}
\frac{E2}{M1}&=&\frac{1}{3} \frac{M^{E2}}{M^{M1}} \\
\frac{C2}{M1}&=&\frac{1}{2\sqrt{2}}
\frac{M^{C2}}{M^{M1}}.
\end{eqnarray}
These ratios (EMR and CMR, respectively) are
equal for $|\bbox{q}|\rightarrow 0$,
a limit
which is never met since, even at the photon point, a finite
$|\bbox{q}|$ is
needed for the transition to take place.

In most calculations reported in the literature
the transition
amplitudes are computed in the rest frame of the $\Delta$.
Such a choice
corresponds to $x=0$ in the expressions of Section \ref{secIII}.
The nucleon four-momentum
$(E_N, -\bbox{q})$ and the photon four-momentum $(\omega,\bbox{q})$
completely specify the
kinematics and the (invariant) photon virtuality, $Q^2=-q^2$ is the
appropriate
quantity in terms of which the electroproduction amplitudes should be
expressed.  In the $\Delta$ reference frame,
\begin{equation}
|\bbox{q}|^2= \left( \frac{m_\Delta^2+m_N^2+Q^2}{2 m_\Delta}
\right)^2- m_N^2
\end{equation}
and
\begin{equation}
\omega= \frac{m_\Delta^2-m_N^2-Q^2}{2 m_\Delta}.
\end{equation}

\begin{figure}
\centerline{\epsfig{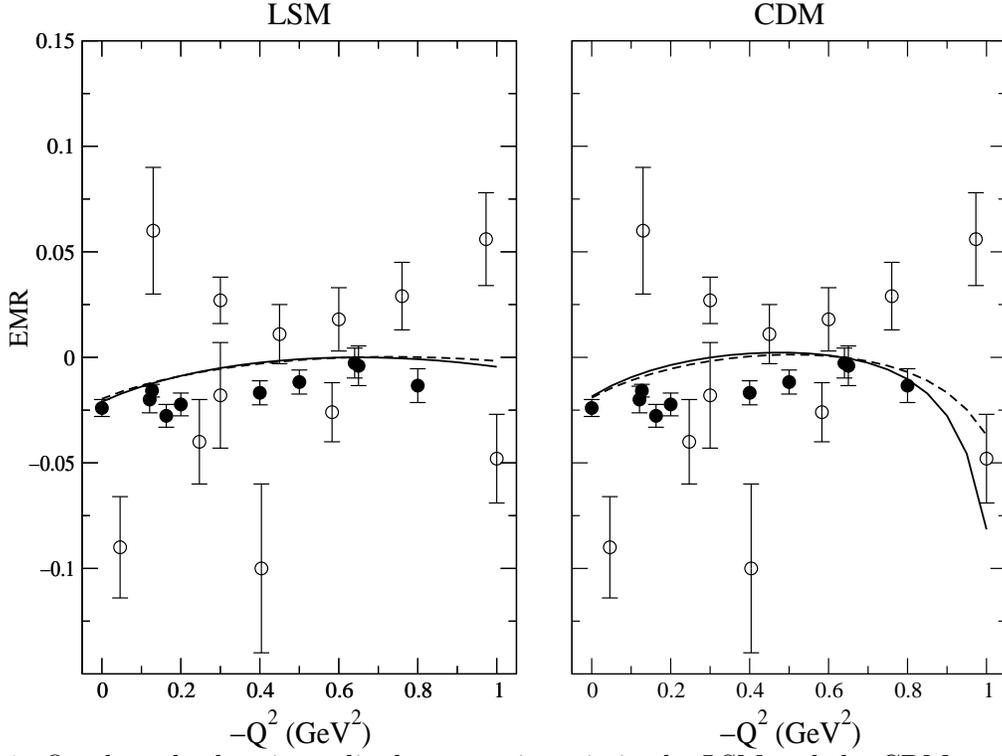}   }
\caption{Quadrupole electric to dipole magnetic ratio in
the LSM and the
CDM, as a function of $-Q^2$. The solid (dashed)
lines show the results
with (without) recoil effects for the standard parameter set
and in the rest frame of the $\Delta$. Experimental data was taken
from~\protect\cite{eold1,eold2} ($\circ$)
and~\protect\cite{enew1,enew2,enew3} ($\bullet$).}\label{fig:emrs}
\end{figure}

\begin{figure}
\centerline{   \epsfig{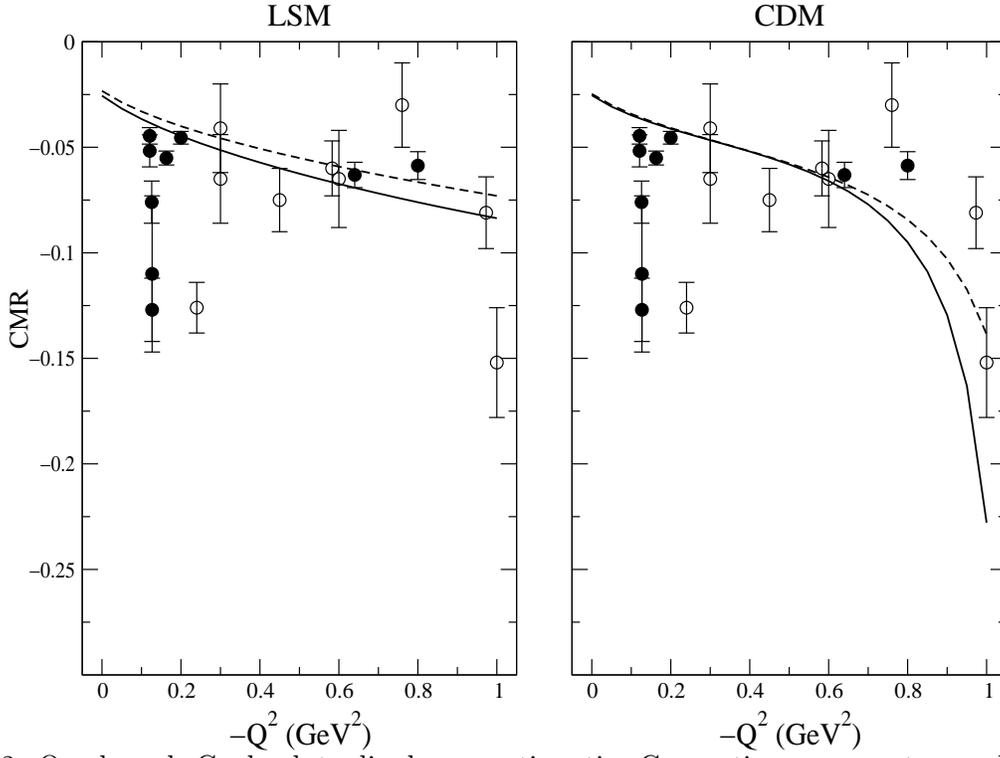}   }
\caption{Quadrupole Coulomb to dipole magnetic ratio.
Conventions, parameters and reference frame as in
Fig.~\ref{fig:emrs}. Experimental data taken
from~\protect\cite{eold1,c2old} ($\circ$)
and~\protect\cite{enew2,c2new} ($\bullet$).  }\label{fig:cmrs}
\end{figure}

\begin{figure}
\centerline{   \epsfig{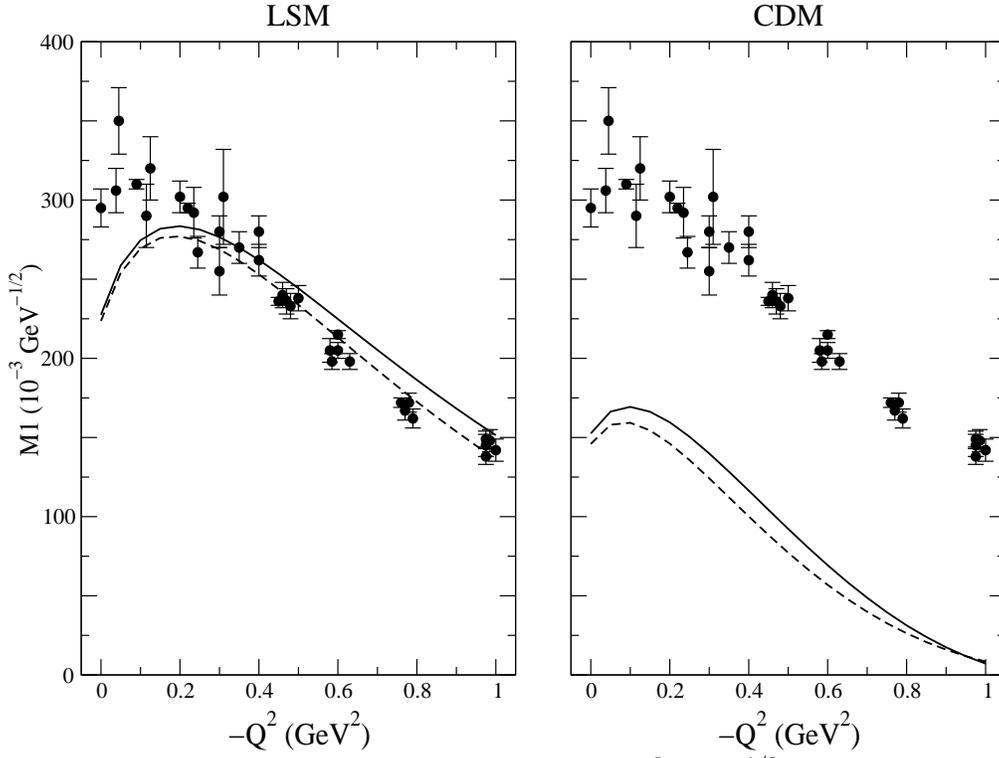}   }
\caption{Dipole magnetic amplitude $M1$ (in units of
10$^{-3}$\,GeV$^{-1/2}$) as
a function of $-Q^2$, for the LSM and the CDM.
Conventions, parameters and reference frame as in
Fig.~\ref{fig:emrs}. Experimental data taken 
from~\protect\cite{m1}.}\label{fig:m1s}
\end{figure}

Figure~\ref{fig:emrs} shows the results for the quadrupole
electric to dipole magnetic ratio, in the LSM and CDM,
 for standard parameter sets in both models,
in the rest frame of the delta.
Figure \ref{fig:cmrs} displays the quadrupole
coulomb to dipole magnetic ratio as a function of $-Q^2$. The
first conclusion to be drawn is the
compatibility of the model predictions with the data,
namely the negative signs for both ratios. From the
theoretical point of view
we don't find any sign of the up and down behaviour
of the data points.
Another interesting conclusion is the small effect of
the recoil corrections
in EMR and CMR. Recoil corrections enhance the nucleon
magnetic moments
\cite{neuber,neuberg} and nucleon magnetic form factors
\cite{drago}. Such
enhancement is also found in the nucleon-delta magnetic
transition as it is
shown in Figure \ref{fig:m1s}, improving the comparison
with experimental values, but the effect, in the present case,
is smaller than for the nucleon.

A similar enhancement turns out to show
up in the quadrupole
electric and coulomb multipoles, and altogether no sensible
modification
appears in EMR and CMR. In the CDM the modification of
$M^{M1}$ due to a
better treatment of the kinematics of the nucleon and
the delta is not enough
to achieve a better comparison with the data. The comparison
with the data of
this observable favours the model with large number of pions
in the cloud. The
big slope of the theoretical CMR in the CDM is due to the small
value predicted for $M^{M1}$ in this model.

The values of the ratios at the photon point
(delta photoproduction) are
$-2.56$\% (LSM) and -2.54\% (CDM) for the CMR and -2.11\% (LSM),
-1.85\% (CDM) for the EMR. These values
are compatible (although slightly smaller, in the case of the EMR) with the
experimental value
$-2.5\pm 0.5$\% estimated for EMR by the Particle
Data Group \cite{pdg}.

\begin{figure}
\centerline{   \epsfig{file=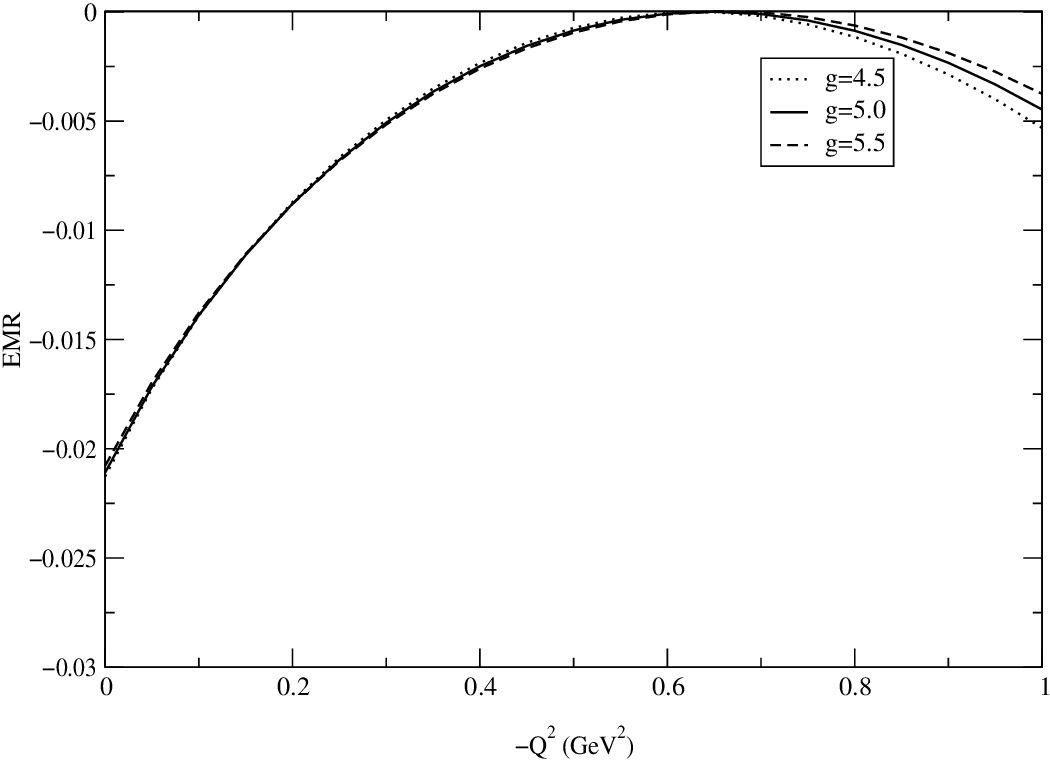,width=70mm,angle=-90}   }
\caption{EMR in the LSM for three values of the coupling constant.}
\label{fig:emrg}
\end{figure}

\begin{figure}
\centerline{    \epsfig{file=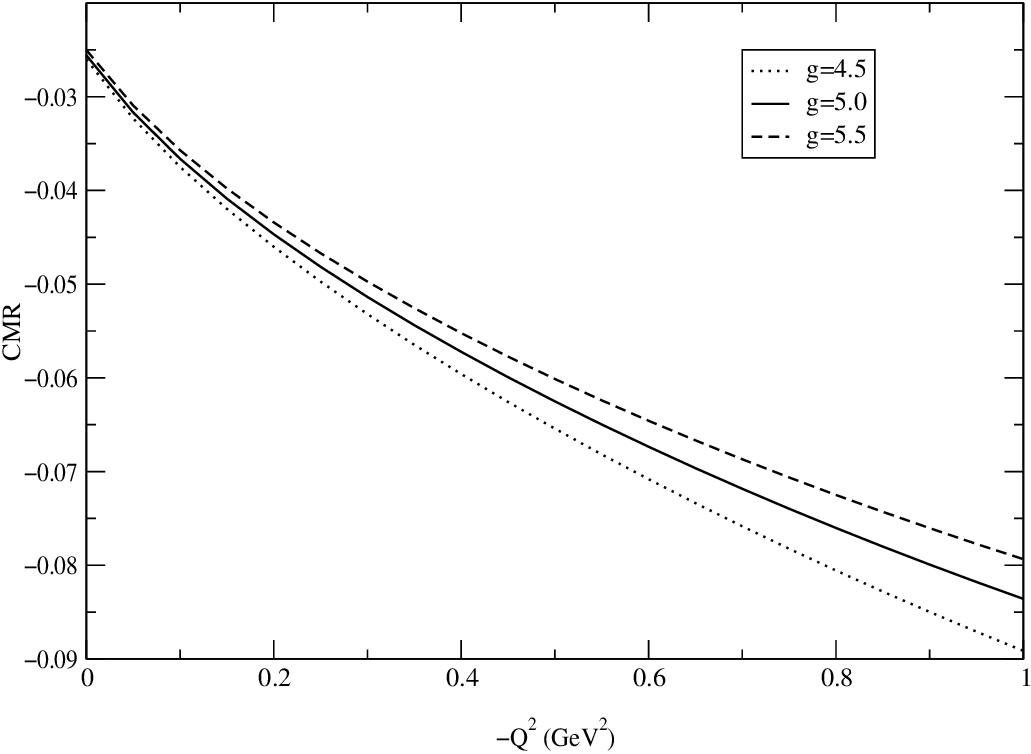,width=70mm,angle=-90}    }
\caption{CMR in the LSM for three values of the coupling constant.}\label{fig:cmrg}
\end{figure}

\begin{figure}
\centerline{    \epsfig{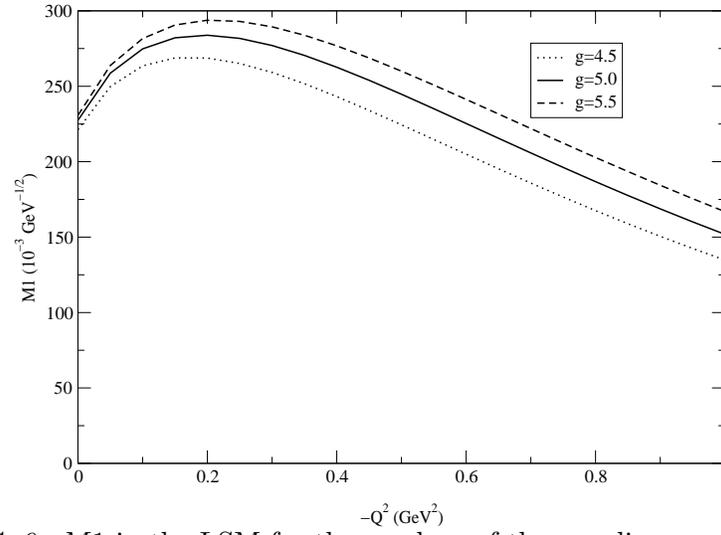}   }
\caption{$M1$ in the LSM for three values of the coupling constant.}
\label{fig:m1g}
\end{figure}

It is not our purpose to find fittings  of model
parameters that better reproduce the experimental results
(model parameters were fixed in the nucleon sector of the models).
Nevertheless it is interesting to analyze the dependence
of the results with model parameters namely
the coupling constants.  As stated before,
the physical window
for $G$ in the CDM is relatively narrow and the resulting
radial wave
functions are very much similar throughout that physical range.
The LSM, on the other hand, provides a larger range and a
large variety of
radial wave functions. The results are summarized in figures
\ref{fig:emrg}--\ref{fig:m1g} for three values of the
 coupling constant in the
LSM: $g=4.5$ (weak coupling, weak pion cloud), $g=5.0$
(intermediate coupling,
standard parameter) and $g=5.5$ (strong coupling,
strong pion cloud).
The graphs correspond to the calculation with recoil
corrections. As Figure \ref{fig:emrg} reveals, the EMR
remains impressively unchanged with $-Q^2$. The CMR
(Figure \ref{fig:cmrg}) is affected specially for large
$-Q^2$. The effect on
$M1$ is shown in Figure~\ref{fig:m1g}. The multipole $C2$
results from pion
contribution alone, whereas $M1$ receives contributions
from both pions and
quarks. The stronger pion cloud enhances more $C2$ than
$M1$ resulting  in a
larger (in absolute value) CMR for higher coupling constant.
The same trends
were also found in the CDM (but with even smaller variations
with $G$).

\begin{figure}
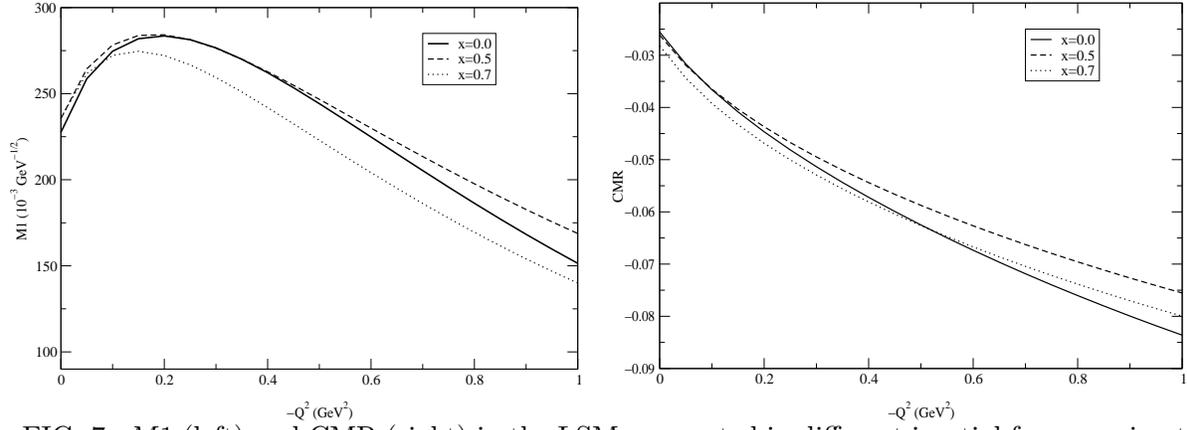

\centerline{   \hfill
   \epsfig{file=fig7a.eps,width=0.34\textwidth,angle=-90}$\hfill$
   \epsfig{file=fig7b.eps,width=0.34\textwidth,angle=-90}
   \hfill
   }
\caption{$M1$ (left) and CMR (right) in the LSM, computed
in different inertial
frames using the standard parameter set in the LSM. The
parameter $x$ denotes the
fraction of the photon momentum carried by the $\Delta$.
}\label{fig:deponx}
\end{figure}

Finally we address the problem of the reference frame.  In principle, the
theoretical amplitudes should not be dependent on the particular choice of the
reference frame. However, due to the lack of translational invariance of the
model baryon states (even when recoil corrections are taken into account), that
is not the case. Nevertheless, no dramatic changes in the results, as a consequence of
the different choice of reference frame, are supposed to occur. In
Figure~\ref{fig:deponx} we present the $M1$ multipole amplitude and the CMR ratio
for the LSM ($g=5.0$) and for three values of the parameter $x$ which is the
fraction of the photon momentum carried by the $\Delta$: $x=0$, $x=0.5$ and
$x=0.7$ (EMR follows the trend of CMR).
The curve
$x=0.5$ would correspond to the Breit frame if nucleon and delta were
degenerate.
The major differences (indicating lack
of covariance) come up at large values of $-Q^2$ as one would anticipate.
Indeed, unlike the correct description of baryon motion through Lorentz boosts
of zero momentum eigenstates as mentioned in Section \ref{secII}, 
our approximate treatment is
{\em not\/} relativistic and, therefore, more reliable for small and intermediate
linear momenta.
The region spanned by the curves in  \ref{fig:deponx} gives an
idea of the ``theoretical uncertainty" of the model predictions.

\section{Conclusions}
\label{secV}

In this paper we addressed the question of the delta electroproduction
amplitudes in the framework of two chiral effective models of the
nucleon with meson and quark degrees of freedom.  Although the
predictions for the ratios $E2/M1$ and $C2/M1$ are compatible with data in
both models, the amplitudes are better reproduced in the LSM thus
favouring a picture of the nucleon and the delta with a stronger pion
cloud.  Recoil corrections of the baryons were taken into account in
this study but no dramatic change was actually found with respect to the
calculation with just angular momentum projection from the hedgehog.
This is different from the modifications occurring in nucleon form
factors where larger effects were found when the center of mass
motion spurious components are removed from the baryon wavefunctions.
Strong fluctuations on $C2/M1$, as seen in the experimental data, are not
observed in the present approach.

\vskip0.5cm

{\em Acknowledgements:}
We thank  B. Golli, M. Rosina, S. \v Sirca, A. Silva and D. Urbano
for useful discussions.
Financial support from Funda\c c\~ao para a Ci\^encia e
Tecnologia, Portugal, project PRAXIS/PCEX/P/FIS/6/96 is
acknowledged. One of the authors (LA)
also acknowledges the financial support to his
PhD program from PRODEP Project No.~185/007.

\appendix
\section{Definition of the magnetic and electric
multipole amplitudes}
\label{appa}
The operator involved in the calculation of the
transverse helicity amplitudes
is
\begin{equation}
\hat{O}(\bbox{r})=e^{-i\bbox{q}\cdot\bbox{r}}
\bbox{\epsilon}_1\cdot
\bbox{J}(\bbox{r}).\label{eq:aa1}
\end{equation}
Choosing the $z$-axis in the direction of $\bbox{q}$, the expansion
of the exponential reads
\begin{equation}
e^{-i\bbox{q}\cdot\bbox{r}} = \sqrt{4\pi}\sum_l (-i)^l\sqrt{2l+1}\,
j_l(qr) Y_{l0}(\hat{\bbox{r}}),
\end{equation}
where $j_l(x)$ are the spherical Bessel functions
and $Y_{lm}(\hat{\bbox{r}})$
are the spherical harmonics. Now, the product
$Y_{l\mu}(\hat{\bbox{r}}) \bbox{\epsilon}_\nu$ can
be cast in terms of the vector
spherical harmonics as
\begin{equation}
Y_{l\mu}(\hat{\bbox{r}}) \bbox{\epsilon}_\nu =
\sum_{jm}\langle l\mu;1\nu
|jm\rangle\,\bbox{Y}^m_{jl}(\hat{\bbox{r}}),
\end{equation}
where $\langle l\mu;l'\mu'|jm\rangle$ are  Clebsch-Gordan
coefficients. Equation~(\ref{eq:aa1}) then reads
\begin{equation}
\hat{O}(\bbox{r})=\sqrt{4\pi}\sum_{lj} (-i)^l \sqrt{2l+1} j_l(qr)
\langle l0;11|j1\rangle \bbox{Y}^1_{jl}(\hat{\bbox{r}})\cdot
\bbox{J}(\bbox{r}).
\end{equation}
Since only the terms with $j=l$, $j=l\pm 1$ contribute, we may write
\begin{eqnarray}
\hat{O}(\bbox{r}) &=& \sqrt{2\pi} \sum_L\sqrt{2L+1} (-i)^L
\left\{ - j_L(qr) \bbox{Y}^1_{LL}(\hat{\bbox{r}})
\cdot\bbox{J}(\bbox{r})
\rule{0mm}{6mm}\right.\nonumber\\
&&+\left.
i\left[\sqrt{\frac{L+1}{2L+1}} j_{L-1}(qr) \bbox{Y}^1_{L\,L-1}
(\hat{\bbox{r}})
\cdot \bbox{J}(\bbox{r})-
\sqrt{\frac{L}{2L+1}} j_{L+1}(qr)\bbox{Y}_{L\,L+1}(\hat{\bbox{r}})
\cdot\bbox{J}(\bbox{r})\right]\right\} \nonumber \\
&=&\sqrt{2\pi} \sum_L\sqrt{2L+1} (-i)^L
\left\{ - j_L(qr) \bbox{Y}^1_{LL}(\hat{\bbox{r}})
+\frac{1}{k}{\bbox{\nabla}\times}\left[j_L(qr)
\bbox{Y}^1_{LL}(\hat{\bbox{r}})\right]
\right\}\cdot\bbox{J}(\bbox{r})
\end{eqnarray}
(see~\cite{eisgrein} for more details). The two terms
inside the curly braces in this expression are respectively
the electromagnetic
field $(L,1)$-magnetic and $(L,1)$-electric multipoles
\begin{eqnarray}
\bbox{A}^{(\mathcal M)}_{LM}(\bbox{r})&=&j_L(qr)
\bbox{Y}^M_{LL}(\hat{\bbox{r}})\\
\bbox{A}^{(\mathcal E)}_{LM}(\bbox{r})&=&-\frac{i}{k}
\bbox{\nabla}\times
\left[j_L(qr)
\bbox{Y}^M_{LL}(\hat{\bbox{r}})\right].
\end{eqnarray}
As it is shown for instance in \cite{eisgrein}, the scalar
products of the $L$-th order field multipoles with
any vector (like the current
density operator) form the irreducible components
of rank-$L$ operators, with
parity $(-1)^{L+1}$ and $(-1)^L$ for the magnetic
and electric multipoles,
respectively. In a transition between states with
angular momentum
$J_i=1/2$ and $J_f=3/2$ and positive parity, only
($L=1$)-magnetic and
($L=2$)-electric multipoles may contribute, so that
the operator $\hat{O}$ may be replaced by
\begin{equation}
\hat{O}'(\bbox{r}) = i\left[
\sqrt{6\pi} \bbox{A}^{(\mathcal M)}_{11}(\bbox{r}) -
\sqrt{10\pi}\bbox{A}^{(\mathcal E)}_{21}(\bbox{r})
\right]\cdot \bbox{J}(\bbox{r}).
\end{equation}
The $M1$ and $E2$ amplitudes are, respectively, the matrix
elements of the first and second terms on the right hand
side of this equation. To make the correspondence with
equation~(\ref{m1def}) we note that
\begin{eqnarray}
i\sqrt{6\pi}\bbox{A}^{(\mathcal M)}_{11}(\bbox{r})
\cdot\bbox{J}(\bbox{r}) &=&
\frac{3i}{\sqrt 2} j_1(qr)\sum_{\mu\nu}\langle 1\mu;1
\nu|11\rangle
\hat{r}_\mu J_\nu(\bbox{r})\nonumber\\
&=&-\frac{3}{2} j_1(qr)\left[\hat{\bbox{r}}\times\bbox{J}
(\bbox{r})\right]_1,
\end{eqnarray}
where use was made of the definition of the vector spherical
harmonics and of
the expression of the spherical components of the vector
product of two
vectors. The scalar amplitude can be derived in a similar fashion.

\section{Transition overlap}
\label{ch:trov}

The transition overlap of two (not necessarily the same)
hedgehog baryons,
defined as
\begin{equation}
N(\bbox{a},\Omega)=\langle\psi'_h|U(\bbox{a})R(\Omega)
|\psi_h\rangle,
\end{equation}
is a recurring function in calculations involving isospin
and linear momentum
projected states \cite{neuber,neuberg}. It is the following
product of quark and meson overlaps:
\begin{displaymath}
N(\bbox{a},\Omega)=N^3_q(\bbox{a},\Omega)\,N_\pi(\bbox{a},\Omega)
N_\sigma(\bbox{a},\Omega)\,N_\chi(\bbox{a},\Omega).
\end{displaymath}
An explicit form for this function can be derived following
the calculation of
the norm overlaps in ref.~\cite{neuber}.
The isoscalar meson overlaps do not depend on $\Omega$ or on
the orientation of
$\bbox{a}$, and taking advantage of the properties of the 
coherent states
(for the sigma field for instance), one obtains
\begin{equation}
N_\sigma(\bbox{a},\Omega)\equiv n_\sigma(a)=
\exp\left\{g^\sigma_0(a)-\pi\int_0^\infty dk\;k^2\,\omega_\sigma(k)
\left[{\eta'}^2(k)+\eta^2(k)\right]\right\},
\end{equation}
where we introduced the functions
\begin{equation}
g^\sigma_l(a)=2\pi\int_0^\infty
dk\;k^2\,\omega_\sigma(k)j_l(ka)\eta'(k)\eta(k)\, .
\label{eq:gsigma}
\end{equation}
A similar function should also be defined for the chromodielectric
field.
The quark overlap is also readily computed, because the
spatial part of the
quark wavefunctions is invariant under isospin rotations
and the spin-isospin
part is invariant under space translations. It is given by
\begin{eqnarray}
N_q(\bbox{a},\Omega)&=&n_q(a){\mathcal N}_q(\Omega),\\
n_q(a)&=&\frac{2}{\pi}\int_0^\infty dk\;k^2\,j_0(ka)
\left[\tilde{u}'(k)\tilde{u}(k)+\tilde{v}'(k)\tilde{v}(k)
\right],\\
{\mathcal N}_q(\Omega)&=&\cos\frac{\beta}{2}\,\cos
\frac{\alpha+\gamma}{2}.
\end{eqnarray}
In these equations, $\tilde{u}$ and $\tilde{v}$ are
Fourier transforms of
the quark profiles, given by
\begin{eqnarray}
\tilde{u}(k)&=&\int_0^\infty dr\;r^2\,j_0(kr)u(r)\\
\tilde{v}(k)&=&\int_0^\infty dr\;r^2\,j_1(kr)v(r).
\end{eqnarray}
For the pion field overlap, we get
\begin{eqnarray}
N_\pi(\bbox{a},\Omega)&=&
\exp\left\{-\pi\int_0^\infty dk\;k^2\omega_\pi(k)
\left[\xi'^2(k)+\xi^2(k)\right]\right\}\;
\exp\left\{\frac{1}{3}\left[g^\pi_0(a)+g^\pi_2(a)\right]
\mbox{Tr}R(\Omega)\right\}
\nonumber\\
&&\mbox{\hspace{10mm}}\exp\left\{g^\pi_2(a)\;\strut
\hat{a}_iR_{ij}(\Omega)\hat{a}_j\right\},
\end{eqnarray}
with $g^\pi_l$ defined as for the sigma (see Eq.~\ref{eq:gsigma}):
\begin{equation}
g^\pi_l(a)=2\pi\int_0^\infty
dk\;k^2\,\omega_\pi(k)j_l(ka)\xi'(k)\xi(k).
\end{equation}

In the reference frame with the $z$-axis along the
axis of the rotation $R$,
the quantity $\hat{a}_iR_{ij}(\Omega)\hat{a}_j$ does
not depend on the
azimuthal angle of vector $\bbox{a}$ and that could be
exploited in order to 
simplify some integrations \cite{neuber}. However,
the orientation of the
$z$-axis has already been fixed along the direction
of the photon momentum.
Therefore, in all integrations over $\bbox{a}$ the
transformation
$\bbox{a}\rightarrow{\mathcal T}^{-1}\bbox{a}$ is
made, where ${\mathcal T}$
is the rotation that aligns the $z$-axis with the
axis of the rotation
$R(\Omega)$, and again advantage can be taken from
the above mentioned
independence of the azimuthal angle.  One gets
\begin{equation}
N_\pi({\mathcal T}^{-1}\bbox{a},\Omega)=n_\pi(a){\mathcal N}_
\pi(a,s,\Omega)\, ,
\end{equation}
where $s=\cos\theta_a$ is the cosine of the
polar angle of $\bbox{a}$, and
\begin{eqnarray}
n_\pi(a)&=&
\exp\left\{g_0^\pi(a)-\pi\int_0^\infty dk\;k^2\omega_\pi(k)
\left[\xi'^2(k)+\xi^2(k)\right]\right\}
\\
{\mathcal N}_\pi(a,s,\Omega)&=&\exp\left\{ 2z(a,s)
\left[\cos^2\frac{\beta}{2}\cos^2\frac{\alpha+\gamma}{2}-1
\right]\right\}\\
z(a,s)&=&\frac{2}{3}\left[g^\pi_0(a)\rule{0mm}{4mm} + P_2(s)g^\pi_2(a)
\right]\, ,
\label{eq:zas}
\end{eqnarray}
$P_2(s)$ being the Legendre polynomial of second degree.

\section{Calculation of matrix elements with projected states}
\label{appb}
Here we present some details regarding the calculation of
matrix elements of
operators between isospin and linear momentum projected states, using
the $C2$ amplitude -- see Eq.~(\ref{c2def}) -- as an example.
The other
amplitudes can be obtained in a similar way (details in
\cite{tese}).

The electromagnetic current density for the two effective
theories considered
in this paper, derived using Noether's theorem, reads
\begin{equation}
J^\mu(\bbox{r}) =\sum_{c=1}^3 \bar{\psi}_{(c)}(\bbox{r})\gamma^\mu
\left(\frac{1}{6}+\frac{1}{2}\tau^{(c)}_0\right) \psi_{(c)}(\bbox{r})
+\left[
\vec{\pi}(\bbox{r})\times \partial^\mu\vec{\pi}(\bbox{r})\right]_0,
\end{equation}
where $c$ is a quark index and the cross
product in the second term is in isospin space. The
current density is the sum of an isoscalar operator
\begin{equation}
S^\mu(\bbox{r})=\frac{1}{6}\sum_{c=1}^3 \bar{\psi}_{(c)}
(\bbox{r})\gamma^\mu
\psi_{(c)}(\bbox{r}),
\end{equation}
which, because of isospin conservation, can not contribute
to $N$-$\Delta$
matrix elements, and the zeroth component of an isovector operator
\begin{equation}
V^\mu_{1,t}(\bbox{r})=\frac{1}{2}\sum_{c=1}^3 \bar{\psi}_{(c)}
(\bbox{r})\gamma^\mu
\tau^{(c)}_t \psi_{(c)}(\bbox{r})
+\left[
\vec{\pi}(\bbox{r})\times \partial^\mu\vec{\pi}(\bbox{r})
\right]_t.
\label{isovec}
\end{equation}
The components of isovector operators commute in a well
defined manner with
the isospin-space rotations involved in the isospin
projectors (see \cite{edm},
for instance) and one can show that, regarding the
expression of the $C2$
amplitude,
\begin{equation}
{\mathcal P}^{3/2^\dag}_{1/2,-1/2} V^0_{1\,0}
(\bbox{r}){\mathcal
P}^{1/2}_{1/2,-1/2}=\sum_t c_t V^0_{1\,t}(\bbox{r})
{\mathcal P}^{1/2}_{-(t+1/2),-1/2},
\end{equation}
with
\begin{equation}
c_t=\sqrt{\frac{2}{3}}\left\langle\frac{1}{2},t+\frac{1}{2};1,-t
\right|\left.\frac{3}{2},
\frac{1}{2}\right\rangle.
\end{equation}
We can then write the $C2$ amplitude as
\begin{equation}
M^{{C2}}(q) =-\frac{\sqrt{10\pi}}{(2\pi)^2}\,N_{N\Delta}
\sum_t c_t \int d^3\bbox{a}\, \int d\Omega\;
{{\mathcal D}^{1/2^*}_{-(1/2+t)\;-1/2}}(\Omega)\;F_t(\bbox{a},\Omega),
\label{eq:c2}
\end{equation}
with
\begin{equation}
F_t(\bbox{a},\Omega)=\int d^3\bbox{r}\;j_2(qr)Y_{20}
(\hat{\bbox{r}})
\langle\psi_h(\Delta)|U^\dag\bigl[(x-1)\bbox{a}\bigr]
V^0_{1t}(\bbox{r})R(\Omega)U(x\bbox{a})|\psi_h(N)\rangle.
\label{deff}
\end{equation}
In this expression, $|\psi_h(N)\rangle$ and $|\psi_h(\Delta)
\rangle$ represent
the nucleon and the delta hedgehogs.

The quark component of the isovector part of the charge
density in Eq.~(\ref{isovec}) cannot contribute here since
the $C2$ is a matrix
element of an $L=2$ operator between $s$-wave quark states.
We are then left only with the pion contribution to the
charge density, and we expand the pion field $\vec{\pi}(\bbox{r})$
and its
canonical conjugate $\vec{P}_\pi(\bbox{r})$ in
plane waves:
\begin{eqnarray}
\pi_i(\bbox{r}) &=& \frac{1}{(2\pi)^{3/2}}
\int d^3\bbox{k}\frac{1}{\sqrt{2\omega_\pi(k)}}\left[
a_i(\bbox{k}) e^{i\bbox{k}\cdot\bbox{r}} +
a_i^\dag(\bbox{k}) e^{-i\bbox{k}\cdot\bbox{r}} \right]\\
{P_\pi}_i(\bbox{r}) &=& \frac{i}{(2\pi)^{3/2}}\int
d^3\bbox{k}\sqrt{\frac{\omega_\pi(k)}{2}}\left[
a_i(\bbox{k}) e^{i\bbox{k}\cdot\bbox{r}} -
a_i^\dag(\bbox{k}) e^{-i\bbox{k}\cdot\bbox{r}} \right].
\end{eqnarray}
The translations and/or isorotations of coherent states still
yield coherent
states (the transformed states are still eigenstates of the annihilation
operators) with shifted and/or isorotated amplitudes. Indeed,
for the pion
field one has
\begin{eqnarray}
a_i(\bbox{k})\;{U}(x\bbox{a})R(\Omega)|\psi_h(N)\rangle &=&
   \sqrt{\frac{\omega_\pi(k)}{2}}\; e^{-ix\bbox{k}\cdot\bbox{a}}\;
   R_{ij}(\Omega)\;
   \xi^{(N)}_j(\bbox{k})\; {U}(x\bbox{a})R(\Omega)|\psi_h(N)\rangle,\\
\langle\psi_h(\Delta)|\;{U}^\dag\bigl[(x-1)\bbox{a}\bigr]\;
   a_i(\bbox{k}) &=&
   \sqrt{\frac{\omega_\pi(k)}{2}}\; e^{i(x-1)\bbox{k}\cdot\bbox{a}}\;
   \xi^{(\Delta)}_i(\bbox{k})\;
   \langle\psi_h(\Delta)|\;{U}^\dag\bigl[(x-1)\bbox{a}\bigr].
\end{eqnarray}

Taking now advantage of the hedgehog shape of
the pion coherent state amplitude, Eq.~(\ref{phedgehog}), we can
write the cartesian components of the functions $F_t(\bbox{a},
\Omega)$ in Eq.~(\ref{deff}) as
\begin{equation}
F_i(\bbox{a},\Omega)=-\frac{i}{2\pi}N(\bbox{a},\Omega)\int d^3\bbox{r}\;
j_2(qr)Y_{20}(\hat{\bbox{r}})\,\epsilon_{ijk}R_{kl}(\Omega)
(r_-)_l(r_+)_j\, 
\alpha(a,r,\hat{\bbox{a}}\cdot\hat{\bbox{r}})\, ,
\label{eq:C12}
\end{equation}
where
\begin{eqnarray}
\bbox{r}_+&=&\bbox{r}+(1-x)\bbox{a},\\
\bbox{r}_-&=&\bbox{r}-x\bbox{a},\\
\alpha(a,r,\hat{\bbox{a}}\cdot\hat{\bbox{r}})&=&
\frac{A_N(r_-)B_\Delta(r_+)+A_\Delta(r_+)B_N(r_-)}{r_- r_+},\\
A_B(r)&=&\int dk\;k^2 j_1(kr)\xi_B(k),\\
B_B(r)&=&\int dk\;k^2 \omega_\pi(k) j_1(kr)\xi_B(k)
\end{eqnarray}
and
\begin{equation}
N(\bbox{a},\Omega)=\langle\psi_h(\Delta)|U(\bbox{a})R(\Omega)
                                          |\psi_h(N)\rangle
\end{equation}
is the transition overlap of the $N$ and the $\Delta$ intrinsic
hedgehogs, derived in Appendix~\ref{ch:trov}.

Expanding the products $(r_-)_l\,(r_+)_j$, the function (\ref{eq:C12})
unfolds in three terms,
\begin{equation}
F_i(\bbox{a},\Omega)=F^{(0)}_i(\bbox{a},\Omega)+F^{(1)}_i
(\bbox{a},\Omega)+  F^{(2)}_i(\bbox{a},\Omega)\,,
\end{equation}
with
\begin{eqnarray}
F^{(0)}_i(\bbox{a},\Omega)&=&\frac{i}{2\pi}N(\bbox{a},\Omega)\,a^2\,
     x(1-x)\,\epsilon_{ijk}R_{kl}(\Omega) \hat{a}_j \hat{a}_l
     \int d^3\bbox{r}\;j_2(qr)Y_{20}(\hat{\bbox{r}})
     \alpha(a,r,\hat{\bbox{a}}\cdot\hat{\bbox{r}})\\
F^{(1)}_i(\bbox{a},\Omega)&=&-\frac{i}{2\pi}N(\bbox{a},\Omega)\,a\,
     \epsilon_{ijk}R_{kl}(\Omega)\nonumber\\
     &&\times
     \int d^3\bbox{r}\;r\,j_2(qr)Y_{20}(\hat{\bbox{r}})
     \bigl[(1-x)\hat{a}_j\,\hat{r}_l-x\hat{a}_l\hat{r}_j\bigr]
     \alpha(a,r,\hat{\bbox{a}}\cdot\hat{\bbox{r}}) \\
F^{(2)}_i(\bbox{a},\Omega)&=&-\frac{i}{2\pi}N(\bbox{a},\Omega)\,
     \epsilon_{ijk}R_{kl}(\Omega)
     \int d^3\bbox{r}\;r^2\,j_2(qr)\,Y_{20}(\hat{\bbox{r}})\,\hat{r}_l
     \hat{r}_j \alpha(a,r,\hat{\bbox{a}}\cdot\hat{\bbox{r}})\, .
\end{eqnarray}
Let us focus on $F^{(0)}_i$. In doing the integration over
$\bbox{r}$ we are
not allowed to pick any particularly convenient orientation
for the vector
$\bbox{a}$, whose components are also integration variables.
Still, it is
possible to perform the integration over the azimuthal angle
of $\bbox{r}$
analytically, yielding
\begin{equation}
F^{(0)}_i(\bbox{a},\Omega)=\frac{i}{4}\sqrt{\frac{5}{\pi}}\;x(1-x)\;
a^2\,
S_{02}(a)\;\epsilon_{ijk}R_{kl}(\Omega)\,N(\bbox{a},\Omega)\;
\left(3\hat{a}^2_{3}-1\right)\hat{a}_j\hat{a}_l,
\end{equation}
where the following functions were introduced
\begin{equation}
S_{nl}(a)=\int_0^\infty
dr\;r^{2+n}\,j_2(qr)\int_{-1}^1du\;P_l(u)\alpha(a,r,u),
\end{equation}
$P_l(u)$ are the Legendre polynomials and $u=\cos\theta_r$ is the
cosine of
the polar angle of $\bbox{r}$. Further analytical refinations of
these $F$
functions are not possible because the function $\alpha(a,r,u)$
depends on the
radial profiles of the meson fields, which are known only numerically.

In order to compute the contribution of the function $F^{(0)}$
to $M^{C2}$, we
replace its spherical components in Eq.~(\ref{eq:c2}).
Making the transformation $\bbox{a}\rightarrow{\mathcal T}^{-1}
\bbox{a}$ (see
Appendix~\ref{ch:trov}) the dependence of the integrand on the
azimuthal
orientation of the vector $\bbox{a}$ is restricted to the terms
$\hat{a}_j\hat{a}_l$ and $\hat{a}_3^2\hat{a}_j\hat{a}_l$,
allowing us to
perform the integration over $\phi_a$ analytically.
The contribution of $F^{(0)}$ to the Coulomb amplitude can
then be written as
\begin{equation}
M^{{C2}^{(0)}}(q)=-\frac{\sqrt{10\pi}}{(2\pi)^2}\,N_{N\Delta}\;
\sum_t c_t
\int_0^\infty da\;a^2\int_{-1}^{1}ds\;\int d\Omega
{\mathcal D}^{1/2^*}_{-(1/2+t)\,-1/2}(\Omega)\;I^{(0)}_{t}
(a,s,\Omega),
\label{eq:mc2}
\end{equation}
where $s=\cos\theta_a$ is the cosine of the polar angle of
$\bbox{a}$ and
\begin{equation}
I^{(0)}_t=\int_0^{2\pi}d\phi_a\;F^{(0)}_t({\mathcal T}^{-1}\bbox{a},
\Omega).
\end{equation}
The integration over the azimuthal angle of $\bbox{a}$ can
now be performed analytically, as the dependence on this
variable only appears in the vector
component products such as $\hat{a}_j\hat{a}_l$, etc. The cartesian
components of the
isovector function $I^{(0)}$ are then
\begin{eqnarray}
I^{(0)}_i(a,s,\Omega)&=&i\sqrt{\frac{5\pi}{4}}\;x(1-x)\;
a^2S_{02}(a)\;
n(a){N}(a,s,\Omega)\nonumber\\
&&\left\{
3P_4(s)\;T_i(\Omega)
-\frac{1}{3}\left[P_0(s)-P_2(s)\right]X_i(\Omega)-P_2(s)W_i(\Omega)
\right.\nonumber\\
&&+
\frac{1}{35}\left[7P_0(s)-10P_2(s)+3P_4(s)\right]
\left[2Y_i(\Omega)+X_i(\Omega)\right]\nonumber\\
&&+\left.\frac{3}{7}\left[P_2(s)-P_4(s)\right]
\left[2Z_i(\Omega)+2V_i(\Omega)+W_i(\Omega)+U_i(\Omega)
\right]\right\},
\end{eqnarray}
where 
\begin{equation}
N(a,s,\Omega)=N({\mathcal T}^{-1}\bbox{a},\Omega)
\end{equation}
(see the Appendix~\ref{ch:trov})
and the following isovectors, functions of the Euler
angles alone, have  been introduced
\begin{eqnarray*}
T_i(\Omega)&=&\epsilon_{ijk}R_{kl}(\Omega)T_{3j}(\Omega)T_{3l}(\Omega)
                     T_{33}^2(\Omega)\\
U_i(\Omega)&=&\epsilon_{ijk}R_{kj}(\Omega)T_{33}^2(\Omega)\\
V_i(\Omega)&=&\epsilon_{ijk}R_{k3}(\Omega)T_{3j}(\Omega)T_{33}(\Omega)\\
W_i(\Omega)&=&\epsilon_{ijk}R_{kl}(\Omega)T_{3j}(\Omega)T_{3l}(\Omega)\\
X_i(\Omega)&=&\epsilon_{ijk}R_{kj}(\Omega)\\
Y_i(\Omega)&=&\epsilon_{i3k}R_{k3}(\Omega)\\
Z_i(\Omega)&=&\epsilon_{i3k}R_{kl}(\Omega)T_{3l}(\Omega)T_{33}(\Omega).
\end{eqnarray*}
In these definitions, $T_{ij}$ are the  matrix elements of
the transformation ${\mathcal T}$.
We now note that $R_{kl}T_{3l}=T^{-1}_{ks}T_{sk'}R_{k'l}T^{-1}_{l3}=
T_{3k}$,
because $T_{sk'}R_{k'l}T^{-1}_{l3}=R^{(z)}_{s3}=\delta_{s3}$ is the
rotation
matrix transformed to the frame in which the $z$-axis
is along the axis of
rotation. Then, we immediately obtain $T_i=0$, $W_i=0$ and
$Z_i=\epsilon_{i3k}T_{3k}T_{33}$. Finally, we reconstruct
the spherical
components of $I^{(0)}$, replace the result in
Eq.~(\ref{eq:mc2}) and, after
the contraction with the Wigner matrices ${\mathcal D}$,
the integration over
the Euler angles can proceed analytically, along the lines
followed in ref.~\cite{neuber}. The final result is
\begin{eqnarray}
M^{{C2}^{(0)}}(q)&=&\frac{\pi}{6\sqrt{2}}\;x(1-x)\;N_{N\Delta}
\nonumber\\
&&
\int_0^\infty da\;a^4\;n(a)\,S_{02}(a)
\left[T_{10}(a)-T_{12}(a)-T_{30}(a)+T_{32}(a)\right],
\end{eqnarray}
with $T_{kl}$ defined by
\begin{equation}
T_{kl}(a)=\int_{-1}^1ds\;\frac{e^{-z}}{z}I_k(z)P_l(s),
\end{equation}
where $I_{k}$ are the modified Bessel functions
of the first kind, depending on the parameter $z$ introduced
in~(\ref{eq:zas}).
Using a similar procedure we computed the two remaining
contributions to the
scalar amplitude. The whole method can be applied to
obtain formulas for the magnetic dipole amplitude \cite{tese}.

\newpage

\newpage

\newpage

\end{document}